\documentclass{article}
\usepackage{arxiv}

\usepackage[utf8]{inputenc} 
\usepackage[T1]{fontenc}    
\usepackage{hyperref}       
\usepackage{url}            
\usepackage{booktabs}       
\usepackage{amsfonts}       
\usepackage{nicefrac}       
\usepackage{microtype}      
\usepackage{lipsum}
\usepackage{graphicx}
\usepackage{amsmath}
\usepackage{algorithm,algpseudocode}
\graphicspath{ {./images/} }

\usepackage{xcolor}
\definecolor{red}{rgb}{0,0,0}
\definecolor{blue}{rgb}{0,0,0}

\title{Centralized and decentralized isolation strategies and their impact on the COVID-19 pandemic dynamics}

\author{
 Alexandru Top\^{i}rceanu\\
  Department of Computer and Information Technology\\
  Politehnica University Timi\c{s}oara\\
  Timi\c{s}oara, 300223, Romania \\
  \texttt{alext@cs.upt.ro} \\
  \And
  Mihai Udrescu\\
  Department of Computer and Information Technology\\
  Politehnica University Timi\c{s}oara\\
  Timi\c{s}oara, 300223, Romania \\
  \texttt{mudrescu@cs.upt.ro} \\
   \AND
  Radu M\u{a}rculescu\\
  Department of Electrical and Computer Engineering\\
  The University of Texas at Austin \\
  Austin, TX 78712, USA \\
  \texttt{radum@utexas.edu} \\
}

\begin{document}
\maketitle
\begin{abstract}
The infectious diseases are spreading due to human interactions enabled by various social networks. Therefore, when a new pathogen such as SARS-CoV-2 causes an outbreak, the non-pharmaceutical isolation policies (\textit{e.g.}, social distancing) are the only possible response to disrupt its spreading. Since there exist several non-pharmaceutical strategies to resort to, we need to assess their efficiency for thwarting the pandemic impact. To this end, we introduce the new epidemic model (SICARS) and compare the centralized (C), decentralized (D), and combined (C+D) social distancing strategies, and analyze their efficiency to control the dynamics of COVID-19 on heterogeneous complex networks. Our analysis shows that the centralized social distancing is necessary to minimize the pandemic spreading. The decentralized policy is insufficient when used alone, but offers the best results when combined with the centralized one. Indeed, the (C+D) policy is the most efficient isolation at mitigating the network superspreaders and reducing the highest node degrees to less than 10\% of their initial values. Our results also indicate that a moderate social distancing, \textit{e.g.,} cutting 50\% of social ties, can reduce the outbreak impact by 47\% for the C isolation, and by 31\% for the D isolation. A stronger social distancing, \textit{e.g.,} cutting 75\% of social ties, can reduce the outbreak by 75\% for the C isolation, by 33\% for the D isolation, and by 87\% for the (C+D) isolation. 
Finally, we study the impact of proactive versus reactive isolation strategies, as well as their delayed enforcement. We find that the reactive response to the pandemic is less efficient, and delaying the adoption of isolation measures by over one month (since the outbreak onset in a region) can have dangerous effects. Thus, our study contributes to an understanding of the COVID-19 pandemic both in space (\textit{i.e.}, network-centric analysis of isolation strategies), and time (\textit{i.e.}, delayed application of isolation strategies). 
We believe our investigations have a high social relevance as they provide insights into understanding how different degrees of social distancing can reduce the peak infection ratio substantially; this can make the COVID-19 pandemic easier to understand and control over an extended period.
\end{abstract}


\section{Introduction}

The incidence of the new Coronavirus disease (COVID-19), caused by the severe acute respiratory syndrome coronavirus 2 (SARS-CoV-2), has seen an exponential rise since the end of 2019, affecting all continents as of March 2020 \cite{world2020coronavirus,cohen2020labs}. The number of cases reported to date is likely an underestimation, owing to the deficiencies in surveillance policy and diagnostic capacity both in high and low-income regions \cite{cohen2020labs}. 
Based on various scientifically relevant criteria, the WHO has declared a global COVID-19 pandemic \cite{world2020coronavirus} in March 2020.  

In the absence of an approved pharmaceutical treatment (or vaccine) and in-depth knowledge of the spreading mechanism, the best strategies against COVID-19 consist of reducing the interactions between susceptible and infected individuals, \textit{e.g.}, through early detection and social distancing \cite{hellewell2020feasibility}. 
Indeed, such non-pharmaceutical interventions (NPIs) turned out to be very effective during previous pandemics \cite{markel2007nonpharmaceutical,hatchett2007public}.

The potential effect of social distancing interventions on the COVID-19 has already been studied in Singapore \cite{lai2020severe}. Indeed, Singapore was among the first regions to report imported cases and has so far succeeded in preventing community spread. 
However, the scale and severity of the Singapore interventions are small in comparison with the measures implemented in China in response to COVID-19. The core Chinese interventions include shutting down schools and workplaces, closing roads and transit systems, canceling public gatherings, and imposing a mandatory quarantine on uninfected people (even those without known exposure to the virus) \cite{kupferschmidt2020china}. Although these actions seem to be working so far, imposing similar restrictions in other countries represents an ongoing challenge. To convince people, governments, and public authorities around the world that such extreme limitations are necessary, we need to back them up with scientific evidence.

Given the dynamics of COVID-19 spreading, we can assess the efficiency of the control measures for this novel pathogen by using mathematical modeling coupled with computer simulations of infectious spread under various scenarios. 
To this end, we propose a new agent-based outbreak model called SICARS (\textit{Susceptible} - \textit{Incubating} - \textit{Contagious} - \textit{Aware} - \textit{Removed} | \textit{Susceptible}), which allows us to assess the impact of the centralized and decentralized isolation strategies on COVID-19 spreading across complex heterogenous networks. Consequently, we run simulations of SICARS and test two fundamentally different strategies, as well as their combined effects:

\begin{enumerate}
\item Centralized (C) strategy, such as the government-imposed lockdown or quarantine; this means social distancing by the \textit{synchronized} removal of a specified ratio of node social ties from the entire social network.
\item Decentralized (D) strategy, such as aware-isolation (DA) and auto-isolation (DI); this means an individual-level social distancing by \textit{asynchronously} removing a specified ratio of personal social ties. More precisely, in DA, the individuals who become aware of their sickness cut the social links in their ego-network. In contrast, in DI, the healthy neighbors of sick individuals isolate themselves from the infected. In both scenarios, the social ties are removed repeatedly (\textit{e.g.}, daily) based on a probability parameter.
\item Hybrid (C+D) strategy, whereby both policies are combined, hence the removal of the social ties involves both centralized and individual-level decision mechanisms. To this end, a fraction of social links are synchronously removed from all nodes in the network, then followed by repeated asynchronous distancing through self-aware isolation of sick nodes (DA), as well as auto-isolation of healthy nodes from their sick neighbors (DI).
\end{enumerate}

In contrast to other network approaches like \cite{van2011gleamviz,pastor2015epidemic,viboud2018rapidd}, SICARS benefits from an additional state we introduce at individual-level, which allows both synchronous and asynchronous isolation to be analyzed, as well as their delayed or progressive application with an increase in severity of isolation.
Our model is inspired by SIR \cite{hethcote2000mathematics}, for which there exist many variations including the SI, SIS, MSIR, SEIR, MSEIRS, SI$_1$I$_2$S, SI$_{1|2}$S, and SI$_1$SI$_2$S models \cite{sahneh2014competitive}. 
A modified SEIR model was previously proposed for the Ebola outbreak of 2014 in Western Africa \cite{diaz2018modified}, consisting of the states specific to Ebola: \textit{Susceptible} - \textit{Exposed} - \textit{Infected} - \textit{Hospitalized} - \textit{Buried} | \textit{Recovered}.
In principle, the SEIR model can be a candidate for capturing the states of an individual in contact with the COVID-19 outbreak. However, SEIR lacks the \textit{Aware} state we introduce in SICARS and which we exploit to study the decentralized isolation strategy. We note that, while both \textit{Contagious} and \textit{Aware} are infectious states (\textit{i.e.}, the disease is transmitted), it is only when a node becomes aware of being infected that it may isolate itself by applying social distancing. The contagious period during which nodes are unaware of their infectiousness (and thus are unable to protect others) is one of the main characteristics making the COVID-19 pandemic so dangerous and hard to control \cite{hellewell2020feasibility}.

We note that there exist several studies based on compartmental models that present timely conclusions on understanding the COVID-19 pandemic spreading.
For instance, Koo \textit{et al.} \cite{koo2020interventions} adapt an existing influenza epidemic simulation model (using data from Singapore) to assess the consequences of social distancing on the transmission dynamics. They find that the intervention strategy combining quarantine, school closure, and workplace distancing seems to be the most effective. Indeed, compared to the no interventions scenario, the combined intervention strategy reduces the estimated number of infections in Singapore by 99.3\%. 

Kucharski \textit{et al.} \cite{kucharski2020early} merge a stochastic transmission model with data on coronavirus cases from Wuhan, China; they estimate the early-stage dynamics of the epidemic and calculate the probability that new cases generate outbreaks in other areas. The main finding is that the median daily reproduction number $R_t$ declined significantly, from 2.35 to 1.05, after the introduction of travel restrictions in Wuhan, on Jan 23, 2020.

Prem \textit{et al.} \cite{prem2020effect} use synthetic contact patterns, adapted to reduced social mixing, and employ SEIR to show that the return to work in China should be further delayed (\textit{i.e.}, by one month). They state that the adoption of physical distancing measures for a long time may reduce the number of infections by more than 92\% by mid-2020.

Hellewell \textit{et al.} \cite{hellewell2020feasibility} study a transmission model tailored to the COVID-19 outbreak and present several interesting findings. First, if there would be no transmission likelihood before symptoms onset, the outbreak containment is secure; unfortunately, this is not the case of COVID-19. Second, when the transmission onset or number of initial cases increases, the controllability of the outbreak decreases significantly. Thus, based on the reproduction number $R_t$ to control the epidemic, 50\% to 90\% of contacts need to be traced and isolated. Third, the critical parameter affecting the controllability of the epidemic is the delay between symptoms onset and start of isolation \cite{hellewell2020feasibility}.

Ferretti \textit{et al.} \cite{ferretti2020quantifying} warn that the current epidemic spread is too fast to be contained by manual contact tracing, and suggest the use of digital tools, like smartphones, to accelerate the contact tracing. Indeed, social behavior plays a vital role in the transmission dynamics of diseases \cite{kucharski2014contribution,reluga2010game}. 

Finally, we note that the idea of social distancing, in various forms, is not new to network science and has been proposed earlier for controlling epidemics \cite{wang2015coupled}. 
Glass \textit{et al.} \cite{glass2006targeted} simulate an influenza model on a community representative of a small town in the United States and show that social distancing would be necessary though school closures and home isolation. For more infectious strains, they found that increased levels of quarantine are needed. 
Valdez \textit{et al.} \cite{valdez2012intermittent} adopt an intermittent social distancing strategy to disturb the epidemic spreading process on different complex network topologies. Specifically, a susceptible individual is allowed to interrupt the contact with an infected neighbor based on a probability value. Using percolation theory, the authors find a critical threshold beyond which the epidemic disappears. 
Reluga \textit{et al.} \cite{reluga2010game} employ game theory to estimate the social behavior of individuals when adopting social distancing during an outbreak. They find that the individual adopts a prophylactic behavior only after the epidemic begins and ceases before the epidemic ends. Additionally, the reproduction number $R_t$ must exceed a certain threshold for individuals to feel that social distancing is worth the effort. 
In contrast, our work studies global, local, and hybrid social distancing strategies to assess their effectiveness in controlling the COVID-19 dynamics.

Taken together, these related studies agree on the efficacy of a generalized quarantine and early detection with isolation strategies. We improve the state-of-the-art with the following contributions:
\begin{itemize}
\item In contrast with the COVID-19 epidemic modeling proposed in \cite{koo2020interventions,kucharski2020early,hellewell2020feasibility}, where isolation is modeled by reducing the size of the susceptible compartment, our network-centric approach targets isolation strategies as (local and global) edge removal mechanisms, hence an emergent and more realistic transmission dynamics.
\item As the differential equations of SEIR do not apply to COVID-19, we use distributed simulation on well-known complex topologies instead of the compartmental models based on random uniform contact networks that are typically used to study epidemics spreading \cite{liu2011epidemic,sahneh2013generalized,sahneh2014competitive,prem2020effect,kucharski2020early}. 
\item Instead of focusing on assessing a specific isolation strategy (\emph{e.g.}, Singapore), our study aims at differentiating the efficiency of centralized (global, government-imposed), decentralized (local, self-imposed), and hybrid isolation while using 
 the SARS-CoV-2 specific biological parameters. 
\item We focus on providing a more accurate quantification of the \textit{impact} of different levels of social distancing, and explore the realistic scenarios of delayed and progressive application of isolation, in the context of the current pandemic.
\end{itemize}

\section{Methods}
\label{sec:methods}

\subsection{Isolation strategies}

We consider a complex interaction network represented as an undirected graph $G=\{N,E\}$, where $N=\{n_i\}$ is the set of nodes and $E=\{e_{ij}\, |\, n_i, n_j \in N \}$ is the set of edges. The nodes represent individuals in the society, while edges represent the social interactions between individuals.

The edge removal ratio $0 \leq e\leq 1$ quantifies the ratio of social ties removed from the network $G$. We define two fundamentally different isolation strategies, based on the underlying principle of removing social ties, as follows:

\paragraph{Centralized (C):} A ratio $e$ of edges are randomly removed  from all nodes $n_i \in N$. This strategy is \textit{synchronous}, meaning that its application happens simultaneously across all network nodes during one simulation step (\textit{e.g.}, day); this translates into a reduction of the number of network edges from $|E|$ to $(1-e)|E|$.

\paragraph{Decentralized (D):} We consider this D strategy in two different variants, namely decentralized \textit{aware-isolation} (DA) and decentralized \textit{auto-isolation} (DI). Both variants are \textit{asynchronous} meaning that they are applied on a node-by-node basis, based on a local decision mechanism. For DA, any node that is aware of being infected will self-isolate itself by repeatedly removing $e$ edges from its vicinity. For a high $e$, an infected node can rapidly become disconnected from the network, thus impeding the pandemic outbreak. For DI, any healthy node that has an infected neighbor will remove the corresponding social tie with a fixed probability of $e$. If $e$ is small, the healthy node will likely need several simulation iterations until successfully removing a specific social tie.

\paragraph{Hybrid (C+D):} We also consider the response to a (C+D) isolation strategy by combining all three strategies C, DI, and DA. More precisely, a centralized strategy is first applied on all nodes $n_i \in N$, by randomly removing a proportion of $e$ edges from each node $n_i$. Then, as on ongoing process, each node $n_i$ that becomes infected and \textit{aware} will apply self-isolation in its vicinity $N_i$, and each \textit{susceptible} neighbor $n_j \in N_i$ will apply isolation from the sick node $n_i$ with probability $e$.

Finally, given the global reaction to the COVID-19 pandemic in early 2020 and its social and economic long-term implications, we notice a delayed enforcement of any isolation strategy in various places. Hence, we also analyze the influence of the response delay ($d=\delta_{Delay}$ days) in applying any of the enumerated strategies.

\subsection{SICARS epidemic model}

\begin{figure}[!hbtp]
\centering
\includegraphics[width=0.85\linewidth]{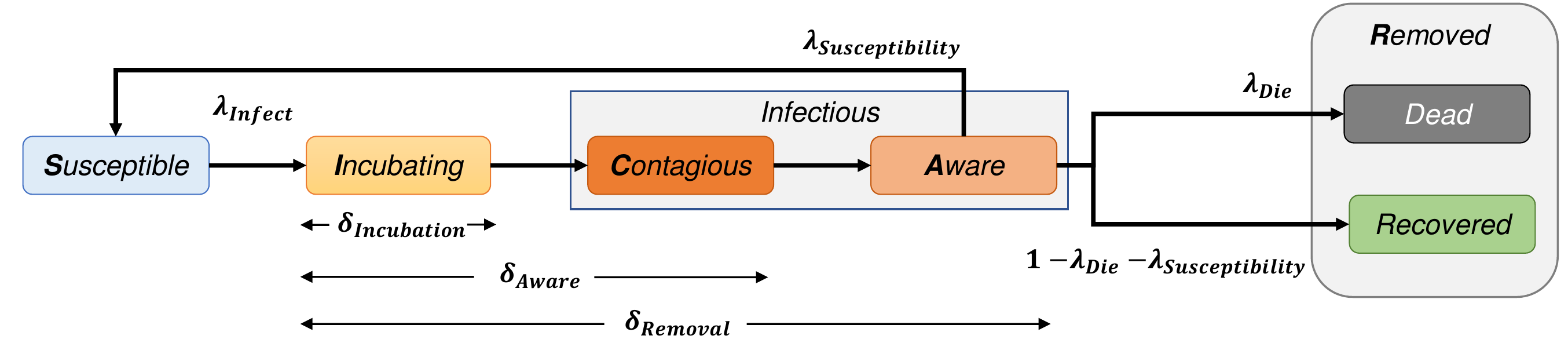}
\caption{The states and parameters defining the SICARS model. By splitting the infectious stage into two sub-states, \textit{contagious} and \textit{aware}, the model becomes unique in its capability of implementing centralized and decentralized isolation strategies.}
\label{fig:states}
\end{figure}

We generalize the popular SIR epidemic model \cite{newman2002spread,pastor2015epidemic} to explicitly enable the analysis of the isolation strategies that are relevant to the COVID-19 outbreak. 
More precisely, our model defines five possible states for an individual: \textit{susceptible} $S$, \textit{incubating} $I$, \textit{contagious} $C$, \textit{aware} $A$, and \textit{removed} $R$, with the caveat that $R$ is equivalent to either \textit{recovered} or \textit{dead}; $R$ can, under specific circumstances, relapse back to the \textit{susceptible} state $S$ (see Figure \ref{fig:states}). We note that SICARS is an \textit{edge removal model}, rather than a node removal model. The reason behind this decision is that removing edges, in the real context of a pandemic, can be controlled easier through isolation measures. In contrast, the removal of nodes means total isolation, a measure that is impractical due to the massive amount of infected patients, and the social consequences of such an extreme isolation.

We assume that, initially, $s$ individuals act as the outbreak seed set $N_s$ (initially marked as \textit{incubating}), while the \textit{susceptible} individuals are the healthy inactive nodes $N \setminus N_s$.
Any \textit{susceptible} node coming in contact with an \textit{incubating} node will not necessarily become infected. On average, only after $\delta_{Incubation}$ days, any \textit{incubating} node automatically becomes \textit{contagious}, and \textit{susceptible} nodes in the vicinity of a \textit{contagious} node become infected with a probability $\lambda_{Infect}$ (see Figure \ref{fig:states}). After some $\delta_{Aware}$ days, the \textit{contagious} nodes become \textit{aware} (\textit{i.e.}, visibly symptomatic), and their infectiousness becomes visible to the neighboring nodes.
From this point on, after a total of $\delta_{Removal}$ days, a node will have transitioned from \textit{incubating} ($I$), to \textit{aware} ($A$), and finally to a \textit{removed} $R$ or \textit{susceptible} $S$ state again. A \textit{removed} node (either \textit{recovered} or \textit{dead}\footnote{We assume merging these two states under the infected nodes gaining immunity assumption, which is likely, but not ultimately confirmed yet in practice.}) will not become \textit{susceptible} again. At the end of the $\delta_{Removal}$ recovery period a node becomes \textit{dead} with probability $\lambda_{Die}$, \textit{susceptible} with probability $\lambda_{Susceptible}$, or \textit{recovered} with probability $1-\lambda_{Die}-\lambda_{Susceptible}$.

\begin{algorithm}
\caption{SICARS model algorithm}\label{sicars-algo}
\begin{algorithmic}[1]
\State \textbf{assign} randomly $n_s \in N_s \leftarrow I$ (\textit{incubating}), where $|N_s| = s$
\State \textbf{assign} all other nodes $n_i \in N \setminus N_s \leftarrow S$ (\textit{susceptible})
\Repeat \Comment{for each iteration day $d$}
\If{$d\geq \delta_{Delay}$}
\State \textbf{if} using (C) strategy: remove any edge $e_{ij} \in E$ with probability $e$
\State  \textbf{if} using (DA) strategy: \textbf{for} $\forall n_k \in A$ (\textit{aware}): remove any adjacent edge $e_{kj} \in E$ with probability $e$
\State \textbf{if} using (DI) strategy: \textbf{for} $\forall n_j \in S$ (\textit{susceptible}), where $n_j \in N_k$ (neighborhood of an \textit{aware} node $n_k$): remove edge $e_{jk}$ (incident to node $n_k$) with probability $e$
\State \textbf{if} using (C+D): apply all corresponding steps (a-c) independently, in this order
\EndIf
\For{$\forall n_k \in A$ (\textit{aware}) for $\geq \delta_{Removal}$ iterations}
\State $n_k \leftarrow R (dead)$ with probability $\lambda_{Die}$
\State $n_k \leftarrow S (susceptible)$ with probability $\lambda_{Susceptible}$
\State $n_k \leftarrow R (removed)$ probability $1-\lambda_{Die}-\lambda_{Susceptible}$
\EndFor
\For{$\forall n_i \in C$ (\textit{contagious})}
\State \textbf{for} $\forall n_j \in S$ (\textit{susceptible}), where $n_j \in N_i$: $n_j \leftarrow I$ (\textit{incubating}) with probability $\lambda_{Infect}$
\State \textbf{if} $n_i \in C$ for $\geq \delta_{Aware}$ iterations: $n_i \leftarrow A$ (\textit{aware})
\EndFor
\For{$\forall n_i \in I$ (\textit{incubating})}
\State \textbf{if} $n_i \in I$ for $\geq \delta_{Incubation}$ iterations: $n_i \leftarrow C$ (\textit{contagious})
\EndFor
\Until {$(|I|=0,|C|=0,|A|=0)$, or ($|R| \geq \theta N (\textit{e.g.},  \theta = 95\%))$}
\end{algorithmic}
\end{algorithm}

\begin{figure}[!hbtp]
\centering
\includegraphics[width=1.0\linewidth]{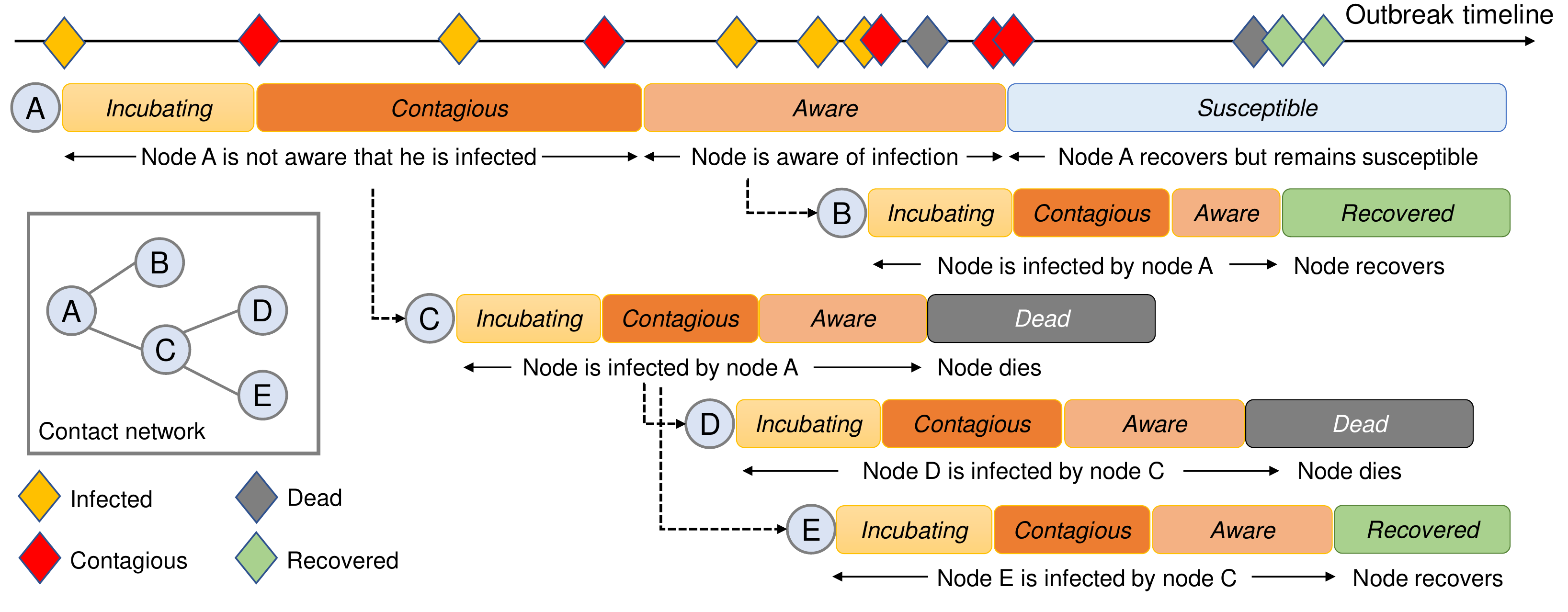}
\caption{Example of an outbreak process according to our SICARS model, when used over a hypothetical contact network with five nodes (A--E), starting with infected node A. After an incubation period, node A becomes contagious and may infect neighbors B and C at any time. Likewise, node C enters the incubation phase and then becomes contagious for nodes D and E. Both contagious and aware states are infectious states, during which a node may spread the disease. Only when a node becomes aware may isolation strategies be applied to it. Finally, all nodes evolve towards one of two states: susceptible or removed (with the mutually exclusive sub-states recovered or dead).}
\label{fig:algorithm}
\end{figure}

The SICARS pandemic simulation based on the model in Figure \ref{fig:states} is formally defined in algorithm \ref{sicars-algo}.
We further illustrate how SICARS works in Figure \ref{fig:algorithm}, with a small example network of five connected nodes, which become infected \cite{hellewell2020feasibility}. The outbreak seed is node A which needs a period of $\delta_{Incubation}$ days to become contagious. After becoming contagious, node A spreads the disease to its neighbors B and C; these, in turn, become incubating after several days of contact with A. Node A is still not aware that it became a disease carrier, nor that it has infected his neighbors. After a delay of $\delta_{Aware}$ days, node A becomes aware of its infectious condition (or state). At this point, nodes B and C know that A is a threat, but are unaware if they have contacted the disease. In the same manner, nodes B and C become contagious, then aware. Nodes D and E are further infected, and the process continues similarly. After $\delta_{Removal}$ days (measured individually for each infected node), every node changes to one of three states: recovered (B, E), dead (C, D), or susceptible (A). The susceptible node A may start the same process all over again, going through all the SICARS states.

\begin{figure}[!hbtp]
\centering
\includegraphics[width=1\linewidth]{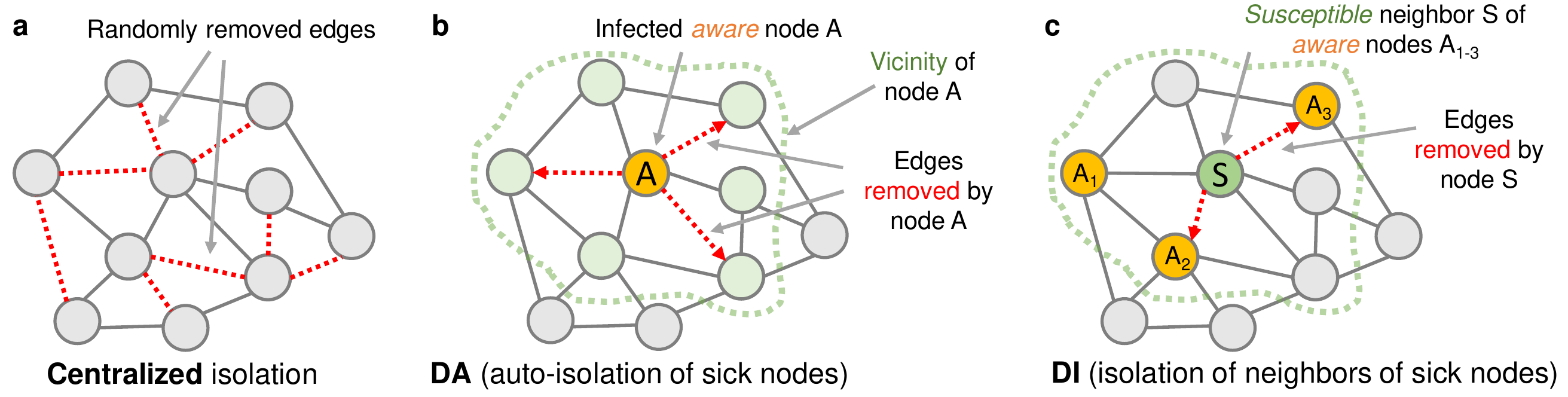}
\caption{Example of applying the C, DA, and DI isolation strategies over a small example contact network using an edge removal ratio $e=0.5$. \textbf{(a)} For C isolation, any edge will be removed with a 50\% probability, regardless of the state of any node. \textbf{(b)} For DA isolation, each node that becomes \textit{aware} will cut all edges in its vicinity with a probability given by $e$; in this example, node A removes 3 out of its 6 adjacent edges. \textit{(c)} For DI isolation, each \textit{susceptible} (healthy) node which detects an \textit{aware} (\textit{i.e.}, sick) neighbor, will try to detach from it with a given probability $e$; in this example, node S manages to detach itself from sick nodes $A_2$ and $A_3$.}
\label{fig:example-cd}
\end{figure}

Furthermore, in Figure \ref{fig:example-cd} we illustrate an intuitive example of applying each isolation strategy independently on a small contact network of ten nodes. By setting $e=0.5$, we show that the C isolation removes about 50\% of edges in the entire network, regardless of the node states (hence, all nodes are colored gray in Figure \ref{fig:example-cd}a); this process is applied \textit{synchronously} over the entire network. The DA and DI isolation, on the other hand, work \textit{asynchronously}, and depend on the (local) development of the disease in the vicinity of each node. In case of DA, pictured in Figure \ref{fig:example-cd}b, only when a node becomes \textit{aware} (orange) can it self-isolate by cutting edges in its vicinity (green nodes). In case of DI, pictured in Figure \ref{fig:example-cd}c, a \textit{susceptible} node (green) can only isolate itself from infected neighbors once these become \textit{aware} (orange), \textit{i.e.}, visibly infected. Finally, in case of the combined C+D strategy, we simply apply C followed by DA and DI independently on the network.

With our SICARS model, we start the diffusion process with $s=10$ initial cases to represent the newly emerging outbreak. The other parameters used in our model are gathered from the recent literature and summarized in Table \ref{tab:parameters}. 

\begin{table}[!h]
\renewcommand{\arraystretch}{1}
\caption{Parameter values used with the outbreak model specific to the COVID-19 pandemic. For \emph{investigated} we experiment with multiple values for the respective parameters.}
\label{tab:parameters}
\centering
\begin{tabular}{lllll}
\hline
Parameter & Empirical value & Model parameter & Value in model & Reference \\
\hline
Incubation period & 5.1-5.8 days & $\delta_{Incubation}$ & 5 days & \cite{lauer2020incubation,backer2020incubation,li2020early} \\
Delay from contact to onset & 11.5-14 days & $\delta_{Aware}$ & 12 days & \cite{lauer2020incubation}\\
Delay from onset to recovery & 2-8 weeks & $\delta_{Removal}$ & 14 days & \cite{eurosurveillance2020updated,mission2020report,linton2020incubation}\\
Death rate & 1-3\% & $\delta_{Die}$ & 2\% & 
\cite{mission2020report,wang2020novel}\\
Susceptibility rate & $<1\%$ & $\delta_{Susceptibility}$ & 1\% & \textit{assumed}\\ \hline
Number of initial cases & unknown & $s$ & 10 & \cite{huang2020clinical}, \textit{assumed} \\
Infectious rate & - & $\delta_{Infect}$ & 0.05 & \cite{pastor2015epidemic}, \textit{assumed} \\
Social tie removal rate & - & $e$ & 0--0.9 & \textit{investigated}\\
\hline
\end{tabular}
\end{table}

\subsection{Test data description}

Our simulations with the SICARS model use several synthetic network topologies and real-world network datasets, \textit{i.e.}, a mesh network ($Mesh$), a Watts-Strogatz small-world network ($SW$), and a scale-free network ($SF$) \cite{wang2003complex}. It is essential to analyze these basic topologies because they possess uniquely distinguishable network properties found in nature \cite{strogatz2001exploring,barabasi2016network}. 
Note that the use of these topologies is \textit{not} an attempt to replace the realistic contact networks, but rather to validate the accuracy and the consistent dynamics of the SICARS model on several different types of graphs.
We also include several sizable real-world network datasets, which present a variety of heterogeneous social interaction networks: the Enron email communication network ($Em$) \cite{leskovec2009community}, the Brightkite location-based social network ($Bk$) \cite{cho2011friendship}, and the Epinions trust network ($Ep$) \cite{richardson2003trust}. 
The reason for choosing these specific datasets is that from a statistics standpoint, they hold characteristics representative for the contact network enabling epidemic spreading. In particular, $Em$ models relevant professional contacts and $Ep$ models relevant friendship contacts. These connections are the most likely to be activated in the real-world social networks. $Bk$ is also an appropriate choice because it models a location-based mesh-like topology, which is a critical aspect of epidemics spreading \cite{smilkov2012influence}.

In Table \ref{tab:datasets}, we present the relevant network measurements for our enumerated test topologies. We include the network size (the numbers of nodes $N$ and edges $E$), average degree $\langle k \rangle$, maximum degree $max(k)$, average path length $APL$, average clustering coefficient $ACC$, network modularity $Mod$, and the network diameter $Dmt$ \cite{wang2003complex}.

\begin{table}[!h]
\renewcommand{\arraystretch}{1}
\caption{The relevant parameters of the synthetic and real-world test networks.}
\label{tab:datasets}
\centering
\begin{tabular}{l|rrrrrrrr}
\hline
$Network$ & $N$ & $E$ & $\langle k \rangle$ & $max(k)$ & $APL$ & $ACC$ & $Mod$ & $Dmt$\\
\hline
$Mesh$ & 5000 & 25234 & 5.047 & 36 & 13.942 & 0.185 & 0.855 & 37\\
$SW$ & 5000 & 19998 & 4.000 & 11 & 10.607 & 0.426 & 0.829 & 22\\
$SF$ & 5000 & 15762 & 3.152 & 294 & 5.378 & 0.007 & 0.640 & 13\\ \hline
$Em$ & 12625 & 20362 & 3.226 & 576 & 3.811 & 0.577 & 0.684 & 9\\
$Bk$ & 58228 & 214078 & 7.353 & 1134 & 7.371 & 0.271 & 0.667 & 18\\
$Ep$ & 75879 & 405739 & 10.694 & 3044 & 11.549 & 0.137 & 0.438 & 14 \\
\hline
\end{tabular}
\end{table}

\subsection{Analytical modeling of SICARS dynamics}

To motivate the importance of simulating SICARS in the context of complex social interaction networks, we describe the dynamics of the model by providing the differential equations that capture the time evolution of the number of \emph{Susceptible}, \emph{Infectious},  and  \emph{Removed} individuals ($S(t)$, $I(t)$, and $R(t)$, respectively). Although relevant for the dynamics of the network topology (which evolves according to the adopted social distancing policies), the number of \emph{Contagious} and \emph{Aware} individuals, $C(t)$ and $A(t)$, are simply a delayed version of $I(t)$. Overall, we have $S(t) + I(t) + C(t) + A(t) + R(t) = N$. 

Another critical parameter that changes over time (as a function of the assumed social distancing policies) is the network density $\rho\left( t\right) = \frac{2\cdot n_{E}\left( t\right)}{N\left( N-1\right)}$, where $n_{E}\left( t\right)$ is the number of edges in the network at moment $t$.
Density $\rho(t)$ has a non-trivial evolution due to the heterogeneous structure of real-world networks, and due to the local decision mechanisms implemented by isolation (\textit{i.e.}, DA, DI); thus, $\rho(t)$ represents a dynamic parameter in our analytical evaluation. 

The equations system describing SICARS is as follows:
\begin{equation}
\begin{cases}
\dfrac{dS(t)}{dt} = -\rho (t)\lambda_{Infect} S(t) (I(t) + C(t) +A (t)) + \lambda_{Susceptibility} A(t) \\
\dfrac{dI(t)}{dt} = \rho (t) \lambda_{Infect} S(t) (I(t) + C(t) +A (t)) \\
C(t) = \int_{t-{\delta}_{Incubation}}^{t} I(\tau ) d\tau \\
A(t) = \int_{t-{\delta}_{Incubation}-{\delta}_{Aware}}^{t} I(\tau ) d\tau \\
\dfrac{dR(t)}{dt} = (1-\lambda_{Susceptibility}) A(t)\\
\end{cases}
\label{eq:sicars_diff}
\end{equation}
with the initial conditions given by:
\begin{equation}
\begin{cases}
S(0) = N-s\\
I(0) = s\\
C(t) = A(t) = R(t) = 0\\
\end{cases}
\label{eq:sicars_diff_init}
\end{equation}

\begin{figure}[!hbtp]
\centering
\includegraphics[width=1.0\linewidth]{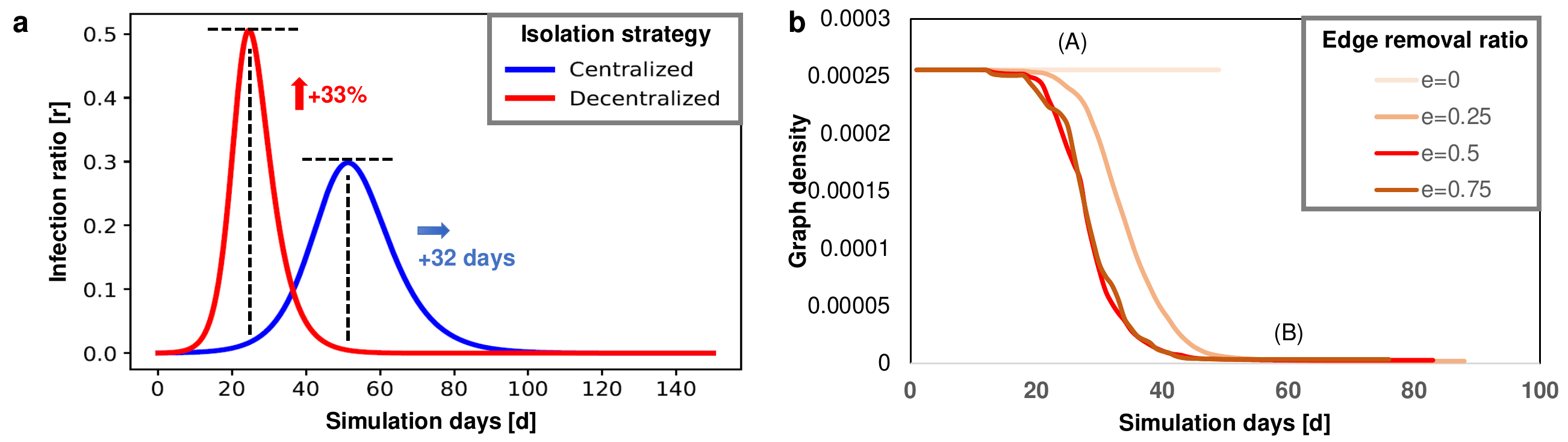}
\caption{\textbf{(a)} Comparison between the C and D social isolation strategies for the analytical SICARS model with equations \ref{eq:sicars_diff} and \ref{eq:sicars_diff_init}. \textbf{(b)} Evolution of network density $\rho(t)$ in time, on the $Em$ network when applying the DA strategy, as a function of the edge removal ratio ($e$).}
\label{fig:math-model}
\end{figure}

The isolation strategies determine the dynamics of $\rho (t)$, which, in turn, interact with the disease dynamics from the equation system \ref{eq:sicars_diff}. In Figure \ref{fig:math-model}a we present the centralized and decentralized evolution of the infection rate ${I(t)}/{N}$ according to the model described in equations \ref{eq:sicars_diff} and \ref{eq:sicars_diff_init}, for an edge removal ratio of $e=0.5$. After applying the C isolation on the $Em$ network (\textit{i.e.}, removing 50\% of network edges), we measure a \textit{constant} density $\rho(t) = 0.0001$. In case of the D isolation, the density evolution is \textit{dynamic} since edge removals are influenced by the spread of disease. We thus obtain the evolution of density on the $Em$ network, for $e=0-0.75$, as shown in Figure \ref{fig:math-model}b; the density evolution (for the same $e=0.5$) is best described by a sigmoid function starting from from $0.00025$ (point A) to $1\cdot 10^{-6}$ (point  B), namely $\rho(t) = 0.00025\left( 1-\frac{1}{1+ e^{-0.3 t + 9}}\right)$.
The analytically determined evolution of the C and D strategies in Figure \ref{fig:math-model}a suggests that the peak infection ratio is higher and achieved sooner in the case of D (red line) compared to C (blue line). Specifically, the infection ratio amplitude is 33\% higher for D and occurs 32 days earlier than for C.


The asymptotic analysis of our model indicates that the centralized policy is more effective than the decentralized one. However, using the network density (\textit{i.e.}, an aggregate parameter) does \textit{not} realistically capture the dynamics of disease spread, since the structure of real-world social interaction networks is not homogeneous but rather heterogeneous in nature. Therefore, to test the interplay between the dynamics of SICARS and the dynamics of the network topology according to the social distancing policies, we use detailed computer simulations as described next.

\section{Results}

\subsection{\textit{How} to control the epidemic dynamics?}

We further analyze how to control the COVID-19 dynamics using social distancing. To this end, we run comprehensive simulations of our SICARS model, for edge removal ratios $e=\{0, 0.1, 0.25, 0.5, 0.65, 0.75, 0.8, 0.9\}$ on all test networks in Table \ref{tab:datasets}; we present next the results corresponding to the C, DA, DI, and C+D isolation strategies. 

Varying the edge removal ratio $e=0-0.9$ enables us to study a broad spectrum of the severity degree of social distancing, \textit{i.e.}, from no distancing at all ($e=0$), to mild ($e=0.1-0.25$), to moderate ($e=0.5-0.65$), all the way to extreme ($e\geq 0.75$) distancing. The effectiveness of these various forms of isolation is relevant to understand more intuitively how exactly social distancing works in many regions of the world. For example, in the USA, a recent poll by Gallup \cite{saad2020spectrum} shows that Americans are progressively increasing their social distancing strictness, from moderate ($e \approx 0.5$, mid-March 2020) to strict ($e \approx 0.75$, end of March 2020). By the end of March 2020, 20\% of polled Americans admit to having no contact with people outside their house, and 44\% of having minimal contact outside their households. Similarly, the strict measures adopted by China \cite{kupferschmidt2020china} and Singapore \cite{lai2020severe} correspond to very strict measures within the social distancing spectrum ($e \geq 0.8$).

\begin{figure}[!hbtp]
\centering
\includegraphics[width=1.0\linewidth]{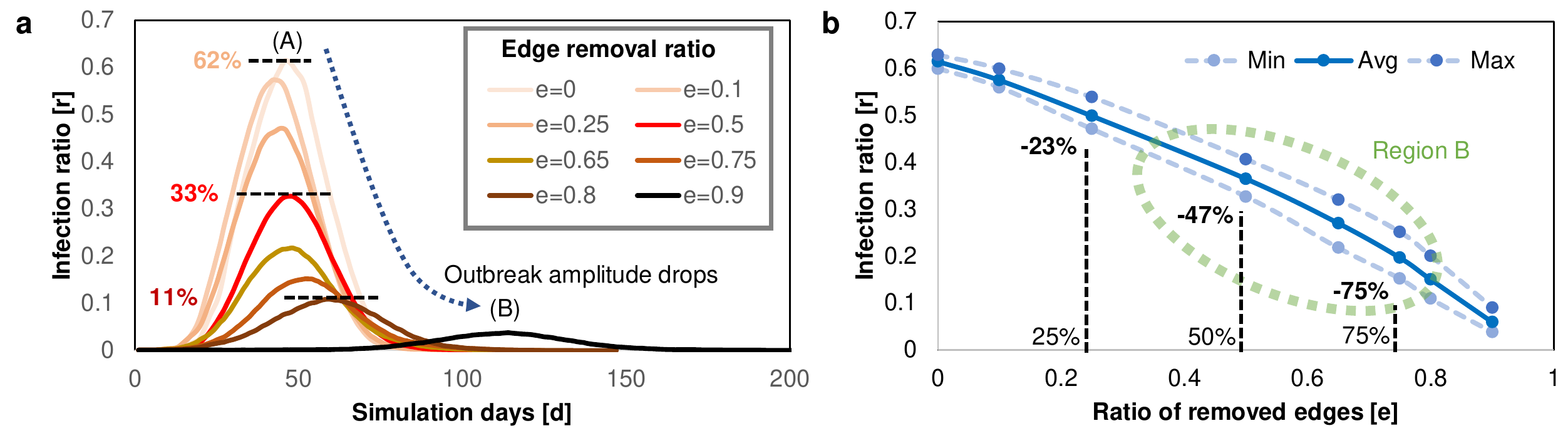}
\caption{Efficiency of the centralized (C) isolation strategy in reducing the ratio of infected individuals. \textbf{(a)} The impact of the edge removal ratio $e$ in reducing and delaying the maximum outbreak amplitude. While no isolation ($e=0$) determines a maximum infection ratio of 62\%, the isolation with $e=0.5$ reduces the peak of the infection ratio to 33\%, and isolation with $e=0.8$ reduces the peak of the infection ratio to just 11\%, but also increases the outbreak duration from roughly 100 days to 180 days. \textbf{(b)} The achieved outbreak amplitude with an increasing $e$, averaged over all test networks in Table \ref{tab:datasets}. We highlight the noticeable linear-concave response of the C strategy in region B (green).}
\label{fig:centralized}
\end{figure}

\paragraph{Centralized isolation:} In Figure \ref{fig:centralized}a, we depict the evolution of the infection ratio ($r = \mbox{number of infected individuals}/ \mbox{population size}$) during the simulation of the SICARS model for the C isolation strategy, using the $Bk$ network as a representative example. Here we measure a drop in the maximum amplitude of the outbreak, from an infection rate of 62\% to just below 4\%, as the ratio of removed edges increases from $e=0$ (\textit{i.e.}, no isolation at all) to $e=0.9$ (\textit{i.e.}, severe isolation with 90\% removed social ties); this means that, for $e=0.9$, we get a 94\% drop in outbreak severity. Additionally, we observe a delay of the maximum amplitude time, from $d=45$ (point A, for $e=0$) to $d=116$ (point B, for $e=0.9$), meaning that the duration of the pandemic is prolonged by about 70 days. In Figure \ref{fig:centralized}b, we display the drop in infection ratio, averaged over all the test networks in Table \ref{tab:datasets}. These results are consistent for all test networks, thus suggesting a significant linear drop in terms of outbreak impact (measured as infection ratio $r$) when a strong centralized social isolation is applied (region B).

To reduce the infection ratio to 50\% of the value corresponding to no isolation at all, we need to remove slightly more than 50\% of social ties. However, if the number of ties is reduced by 80\% ($e=0.8$), we observe a reduction of 82\% in the amplitude of the outbreak.

\begin{figure}[!hbtp]
\centering
\includegraphics[width=1.0\linewidth]{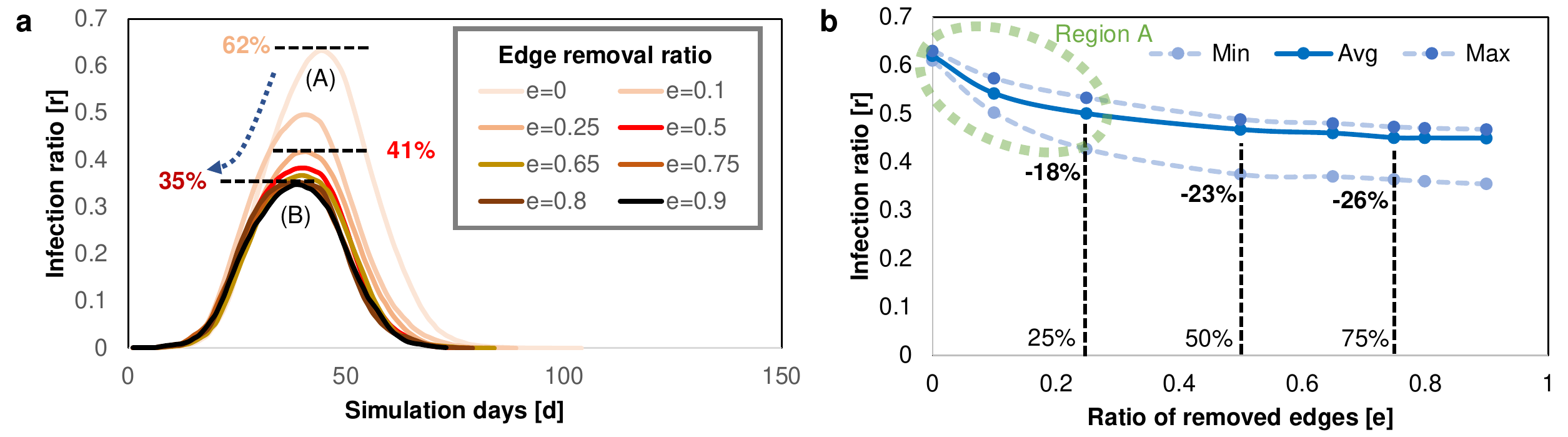}
\caption{Efficiency of the decentralized (D) isolation in reducing the ratio of infected individuals. \textbf{(a)} The impact of the edge removal ratio $e$ in reducing and advancing the maximum outbreak amplitude. While no isolation ($e=0$) determines a maximum infection ratio of 62\%, the isolation with $e=0.5$ reduces the peak infection ratio to 41\%, and isolation with $e=0.8$ reduces the peak infection ratio to 35\%. The outbreak duration is not affected in a significant way. \textbf{(b)} The achieved outbreak amplitude with increasing $e$, averaged over all test networks in Table \ref{tab:datasets}. We highlight the noticeable logarithmic-convex response of the D strategy in region A (green).}
\label{fig:decentralized}
\end{figure}

\paragraph{Decentralized isolation:} We run simulations for both DA and DI on all test networks and find only marginal differences between the two decentralized isolation strategies. Generally, DI follows the same trend as DA on all test networks, but its effects are slightly delayed on time (OX axis) by 0-5\%. 
Figure \ref{fig:decentralized}a depicts the evolution of the infected ratio for the decentralized DA isolation strategy, using the $Bk$ network as a representative example.
By increasing the edge removal ratio from $e=0$ to $e=0.9$, we observe a drop in the maximum amplitude of the outbreak from 62\% to 35\%; this translates into a drop of only 43\% in outbreak severity. Additionally, we observe a slight acceleration of the maximum amplitude time, from $d=46$ (for $e=0$) to $d=39$ (for $e=0.9$), meaning that the duration of the pandemic is shortened by approximately 7 days. In Figure \ref{fig:decentralized}b, we display the drop in infection ratio averaged over all the test networks. The results over the entire set of test networks suggest a moderate logarithmic drop in outbreak impact, even for a strong decentralized isolation (region A).

Reducing the number of social ties has an immediate impact, as a reduction of only 25\% ties can reduce the impact of the outbreak by 18\%; however, further reducing the number of ties renders insignificant improvements. Our results suggest that, by using a decentralized isolation strategy, it is not possible to reduce the infection ratio by 50\% compared to no isolation at all; the best we can achieve is a reduction of about 26\%.

\begin{figure}[!hbtp]
\centering
\includegraphics[width=1.0\linewidth]{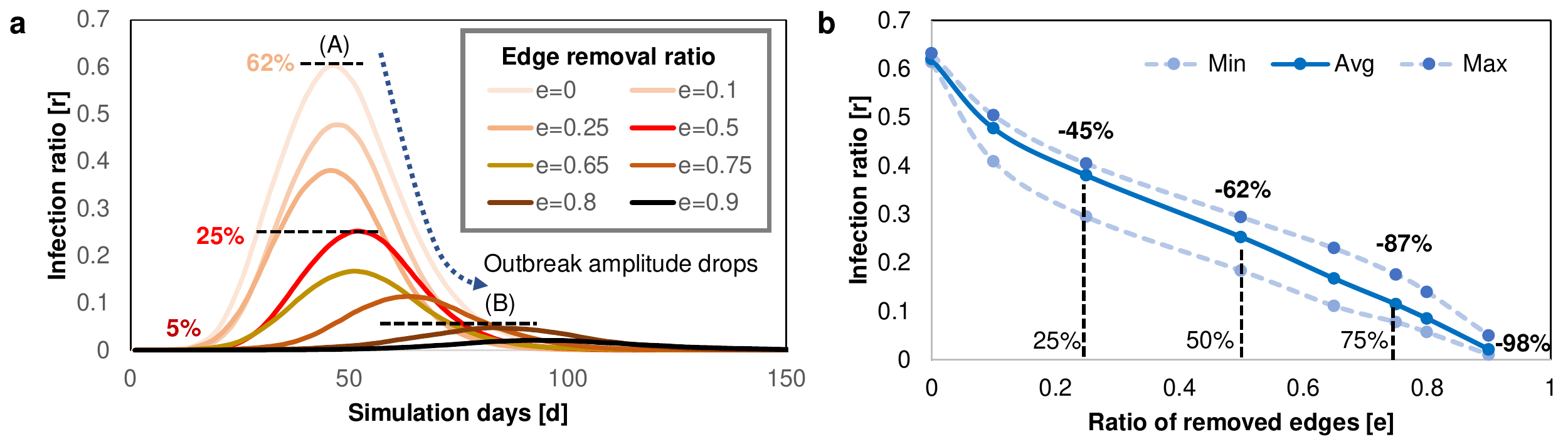}
\caption{Efficiency of the combined (C+D) isolation strategy in reducing the ratio of infected individuals. \textbf{(a)} The Impact of the edge removal ratio $e$ in reducing and delaying the maximum outbreak amplitude. While no isolation ($e=0$) determines a maximum infection ratio of 62\%, the isolation with $e=0.5$ reduces the peak of the infection ratio significantly to 25\%; isolation with $e=0.8$ further reduces the peak infection ratio to just 5\%. However, the outbreak duration increases from roughly 100 days to 140 days. \textbf{(b)} The achieved outbreak amplitude with the increasing $e$, averaged over all test networks in Table \ref{tab:datasets}. Here, we notice both the initial logarithmic-convex signature of the D strategy, as well as the final linear-concave signature of the C strategy combined.}
\label{fig:cd-combined}
\end{figure}

\paragraph{Combined C+D:} 
In comparison to C or D isolation alone, the hybrid C+D gets the best of each strategy (Figure \ref{fig:cd-combined}b). We notice that the maximum infection amplitude has a linear-concave response when using the C isolation (Figure \ref{fig:centralized}b), and a logarithmic-convex response when using the D isolation (Figure \ref{fig:decentralized}b). This difference means that the effects of the D isolation are seen immediately for small increases in $e$ (region A in Figure \ref{fig:decentralized}b). In contrast, moderate to significant increases of $e$ is needed to gain the benefit of C isolation (region B in Figure \ref{fig:centralized}b). In this sense, the C+D strategy shows the same convexity as D, but with a sustained linear reduction that is specific to C.

Figure \ref{fig:cd-combined}a depicts the evolution of the infection ratio $r$ for the combined C+D isolation strategy, using the same $Bk$ network as a representative example. By increasing the edge removal ratio from $e=0-0.9$, we observe a significant drop in the maximum amplitude of the outbreak from 62\% to just 5\% ($e=0.8$), respectively 2\% ($e=0.9$). Compared to C and D isolation, with C+D we measure the highest efficiency in reducing the maximum infection ratio, by as much as 87\% ($e=0.75$), and 98\% ($e=0.9$).
Additionally, we observe a delay of the maximum amplitude time, from $d=45$ (point A, for $e=0$) to $d=70$ (point B, for $e=0.8$), meaning that the duration of the pandemic is prolonged by about 25 days (55\%).

For the hybrid C+D strategy, we need to remove just 25-50\% of social ties in order to reduce the infection ratio to 50\%, compared to no isolation at all. Moreover, if we reduce the number of ties by 75\%, we obtain a significant reduction of 87\% in the outbreak amplitude (Figure \ref{fig:cd-combined}b).



\subsection{\textit{Why} isolation strategies work?}

After analyzing \textit{how} to control the outbreak dynamics, we further try to explain \textit{why} the presented isolation methods work. To this end, we investigate a crucial property of network structure, namely the distribution of superspreaders \cite{pei2014searching} before and after applying any of the discussed strategies. Indeed, many real-world networks possess a power-law distribution of nodes degree, thus favoring the formation of hubs \cite{barabasi1999emergence,barabasi2016network}. These hubs have a significant accelerating effect on the diffusion of innovation, ideas, as well as epidemics \cite{pastor2015epidemic,pei2014searching,lloyd2005superspreading}; this is why these hubs are called superspreaders. During an outbreak, it becomes paramount to reduce the connectivity of these superspreaders because a single infected hub can exceed the combined adverse effects of many average or low connected nodes.  

\begin{figure}[!hbtp]
\centering
\includegraphics[width=1.0\linewidth]{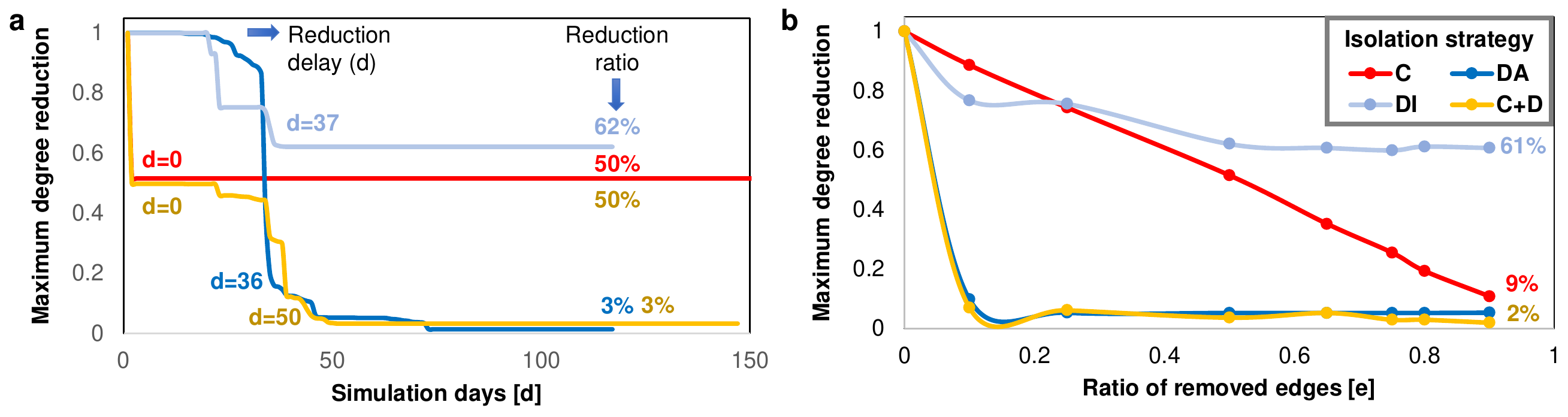}
\caption{Efficiency of the centralized and decentralized strategies in reducing the maximum degree of the network. \textbf{(a)} The reduction delay (horizontal time axis, expressed in days $d$) and reduction ratio (vertical axis); panel {\bf a} is a snapshot of panel {\bf b} specifically for $e=0.5$. Each strategy presents a unique signature: C isolation reduces the maximum degree moderately from the beginning, DA isolation has a strong but delayed reduction, DI isolation has a moderate and delayed reduction, and hybrid (C+D) isolation has an initial moderate reduction, followed by a delayed strong reduction of maximum degree. \textbf{(b)} The effect of the edge removal ratio $e$ on the reduction of maximum degree for each isolation strategy.}
\label{fig:superspreaders}
\end{figure}

The simulation results in Figure \ref{fig:superspreaders}a depict a representative evolution of the maximum degree reduction, by employing the centralized and decentralized isolation strategies for $e=0.5$. Based on their ability to mitigate superspreaders, we notice a new signature for each strategy, which explains why they are more or less efficient at reducing the outbreak impact.
Specifically, the C strategy (red line) reduces the maximum degree to 50\% of the initial network's maximum degree (\textit{i.e.}, given by $e=0.5$) from the very beginning ($d=0$) of the simulation. The DA and DI isolation strategies (blue lines) have a delay of $\approx d=36-37$ days until the network responds; after this moment, DI only reduces the maximum degree to about 62\%, while DA reduces it significantly, to only 3\%. Finally, the hybrid C+D isolation carries a signature that indicates the highest efficiency; namely, the reduction to 50\% is obtained at $d=0$, followed by a reduction to 3\%, similar to the DA strategy delay. The observations discussed for $e=0.5$, in Figure \ref{fig:superspreaders}a, are enforced in Figure \ref{fig:superspreaders}b, as we provide an extended overview of the maximum degree reduction for all tested edge removal ratios $e$. 

Our experimental results support the idea that the DA and C+D strategies ensure that no superspreaders remain in the network when applying an edge removal ratio $e \geq 0.5$. For example, on the $Bk$ network, the C+D strategy reduces the maximum degree from $k=1134$ to $k=34$; on the $SF$ network, the C+D strategy reduces the maximum degree from $k=294$ to $k=16$, thus limiting the impact of hubs in the process of epidemic diffusion.

\subsection{\textit{What} may alter the course of the pandemic?}

\paragraph{Reinfection impact:} An additional parameter worth investigating for our epidemic model is the susceptibility rate $\lambda_{Susceptibility}$, which is still uncertain for the SARS-CoV-2 virus at the end of March 2020. Throughout our simulations, we consider a susceptibility rate of $\lambda_{Susceptibility}=0.01$. However, we further explore the impact of isolation strategies if the recovered patients prove to have a much higher chance of becoming reinfected (up to 10-50\%).

\begin{figure}[!hbtp]
\centering
\includegraphics[width=1.0\linewidth]{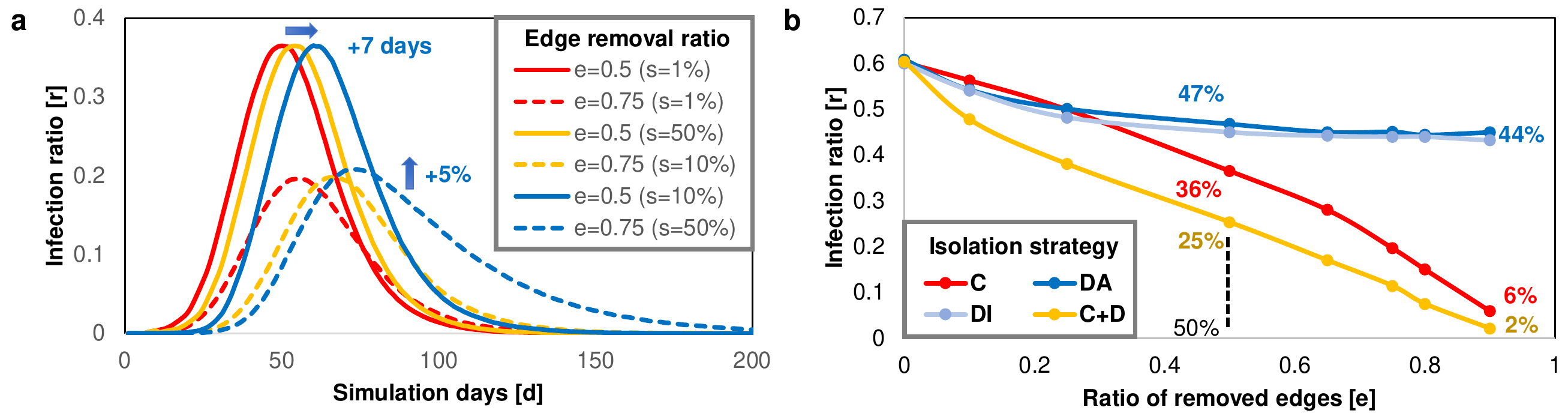}
\caption{The effect of a higher patient relapse rate ($s=1-50\%$) in terms of infection ratio. \textbf{(a)} The effect of different relapse rates when using the C isolation strategy. A higher relapse rate of $s=50\%$ increases the infection ratio by 5\% and delays it by 7 days. \textbf{(b)} The effect of different isolation strategies when using a relapse rate of $s=50\%$. The combined logarithmic-convex and linear-concave of the (C+D) hybrid strategy (yellow) explains its higher effectiveness.}
\label{fig:susceptibility}
\end{figure}

Figure \ref{fig:susceptibility}a shows that the pandemic dynamics is not significantly different at higher susceptibility rates. The moment of the peak infection rate may be delayed by up to one week, while the amplitude of the infection ratio may increase by up to 5\%. Note that these results are based solely on the combination of COVID-19 specific parameters, and may not be directly generalized to other past or future epidemic diseases.

In Figure \ref{fig:susceptibility}b, we focus on a single relapse rate of $s=50\%$ to show the impact of each isolation strategy; the response of all isolation strategies is the same regardless of the relapse rate. The higher efficiency of the C+D isolation is again underlined in panel \ref{fig:susceptibility}b. This set of experiments indicates that even a high patient relapse rate would not alter the effectiveness of the isolation strategies significantly.

\paragraph{Delaying isolation restrictions:} 
While the WHO praised China's quick and aggressive response \cite{kupferschmidt2020china}, other countries, such as the UK or USA, have delayed their responses. We analyze next the efficiency of all isolation strategies in the context of a delayed implementation of quarantines/lockdowns. To this end, we compare three scenarios when the same strategies are applied with a delay of $d=\{0,20,30,50\}$ days after the initial outbreak.

\begin{figure}[!hbtp]
\centering
\includegraphics[width=1.0\linewidth]{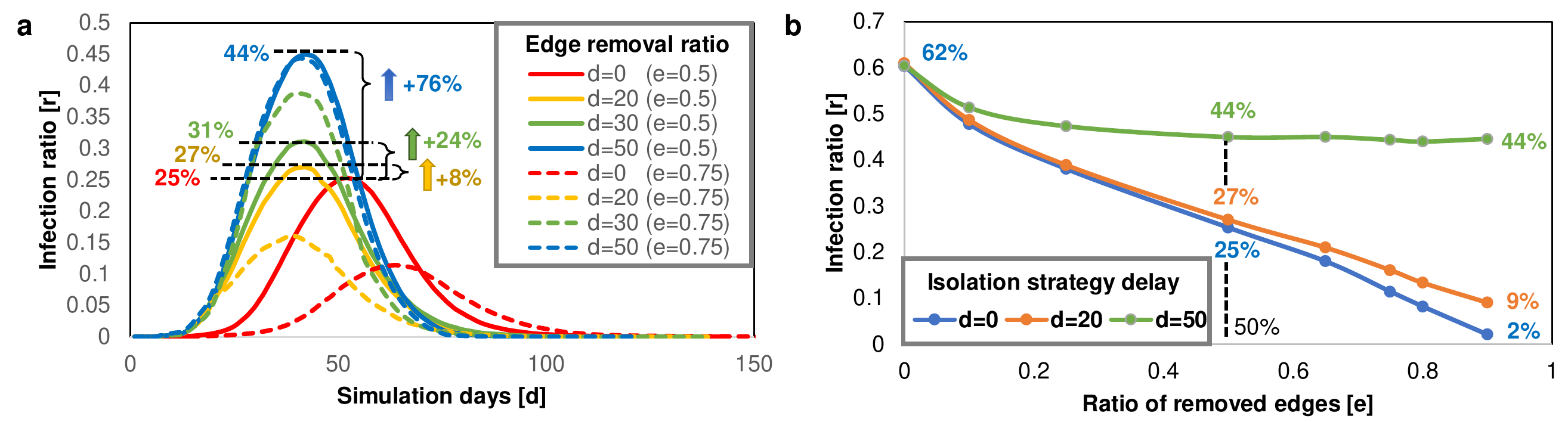}
\caption{Effects of delaying the application of isolation strategies, relative to the start of the outbreak, measured as the maximum infection ratio in the network. \textbf{(a)} Detailed comparison on the outbreak amplitude obtained for delays of $d=\{0,20,30,50\}$ days, with $e=0.5$ and $e=0.75$. For $e=0.5$, a delay of $d=20$ days increases the maximum infection ratio by 8\%; a delay of $d=30$ increases the ratio by 24\%; a delay of $d=50$ increases the maximum ratio by 76\%. \textbf{(b)} Overview of the delays' impact on the maximum infection ratio with increasing values of $e$. The smaller negative impact of the shorter delay ($d=20$, orange line) is distinguishable from the impact of the longer delay ($d=50$, green line).}
\label{fig:delay}
\end{figure}

The simulation results in Figure \ref{fig:delay}a suggest that governments/authorities should take an \textit{immediate action} to enforce the isolation strategies. In general, for moderate isolation ($e=0.5$), the outbreak amplitude increases significantly with the delay (Figure \ref{fig:delay}a); thus, for a delay of $d=20$ days (yellow line) we observe a +8\% higher infection ratio, for $d=30$ days (green line) we observe a +24\% higher infection ratio , and for $d=50$ (blue line) the results are alarming with +76\% more infections than the immediate response (red line). A more severe isolation ($e=0.75$) further amplifies the effects of the delaying isolation strategies, compared to the moderate isolation. As such, for the same delays, we measure increases in the infection ratio of +41\% (yellow dotted line), $3.4 \times$ (green dotted line), respectively $3.88 \times$ higher (blue dotted line) compared to an immediate response (red dotted line). Figure \ref{fig:delay}b presents an overview of each delay impact on the maximum infection ratio for the entire range of tested edge removal ratios $e=0-0.9$. The difference between the shorter delay of $d=20$ days (orange line) and the long delay of $d=50$ days (green line) is noticeable.


\paragraph{Proactive vs. reactive isolation:} Complementary to a delayed application of isolation restrictions discussed in the previous section, we also investigate the feasibility of a reactive policy that is being used by some authorities around the world. In other words, some governments are \textit{reacting} to the accelerated outbreak spread by adopting increasingly more severe quarantine measures as time goes by.
In contrast to the \textit{proactive} isolation (meaning that a fixed $e$ is imposed from a specific moment in time), we model the reactive policy by progressively increasing the strength of isolation, from $e=0$ to $e=0.8$, at fixed moments in time. As such, we analyze two realistic scenarios for adopting a progressive policy. First, the edge removal ratio is increased as a delayed reaction to the pandemic spreading, with values $e=\{0,0.2,0.4,0.6,0.8\}$, applied at moments $r_1=\{10,20,40,60\}$ (faster reaction). Second, the reaction happens at moments $r_2=\{20,40,60,80\}$ (slower reaction). This means that, for example, during the first 20 days of the $r_2$ progressive policy, we have no isolation at all (\textit{i.e.}, $e=0$); after day $d=20$, $e$ is increased to $e=0.2$, and so on. 

\begin{figure}[!hbtp]
\centering
\includegraphics[width=1.0\linewidth]{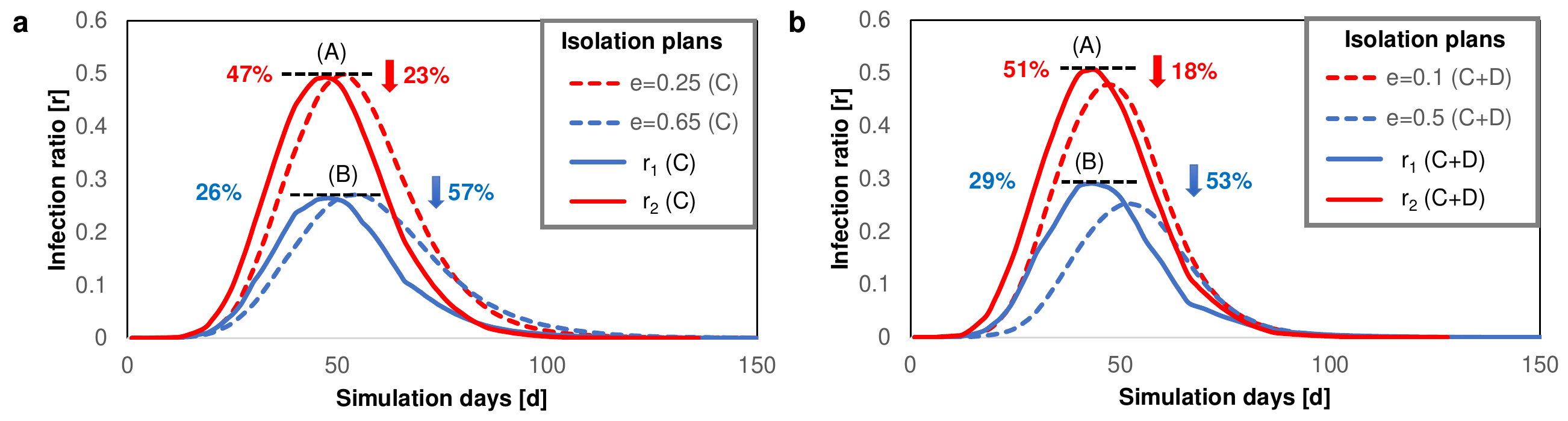}
\caption{Comparison between proactive and reactive responses in applying isolation measured as the maximum achieved infection ratio. \textbf{(a)} C isolation strategies (proactive) with $e=0.25$ (mild isolation) and $e=0.65$ (stricter isolation) are equivalent to reactive policies $r_2$ (slower reaction, every 20 days) and $r_1$ (faster reaction, every 10 days), respectively. \textbf{(b)} C+D isolation (proactive) with $e=0.1$ (very mild isolation) and $e=0.5$ (moderate isolation) are equivalent to reactive policies $r_2$ and $r_1$, respectively. The arrows pointing down suggest the reduction in infection ratio compared to the case of no isolation.}
\label{fig:reactive}
\end{figure}


In Figure \ref{fig:reactive}a we compare the reactive policies $r_1$ and $r_2$ with roughly equivalent proactive C strategies. In this sense, we find that the $r_1$ policies has an equivalent effect in reducing the impact of the outbreak, as the C isolation with $e=0.65$ (point B). Additionally, the $r_2$ policies is similar to C isolation with $e=0.25$ (point A). In other words, if a government reacts slowly, by progressive increases of isolation severity (\textit{e.g.}, policy $r_2$) up to $e=0.8$, the result will be the same as if the government would have applied a more relaxed C isolation with $e=0.25$ from the very beginning of the outbreak.

In the case of the combined C+D strategy, the impact of the progressive isolation is less effective than in the centralized context. To that end, Figure \ref{fig:reactive}b shows that policy $r_1$ is roughly similar to C+D isolation (proactive) applied with $e=0.5$ (point B), respectively policy $r_2$ is similar to C+D isolation with $e=0.1$ (point A). This means that, even though $r_2$ eventually reaches a very strict isolation of $e=0.8$ (after 80 days), it is not more effective than a very mild C+D isolation with $e=0.1$.

\subsection{The impact of isolation strategies on small-scale communities}

While real contact networks have been studied before in the context of epidemic spreading \cite{salathe2010high,genois2018can,stopczynski2014measuring}, their corresponding topologies are representative only for small-scale communities, or specific micro-environments (\textit{e.g.}, schools, stadiums, airports, hospitals).  However, we cannot directly extrapolate the topological characteristics of such contact networks at the level of a metropolis, region, or country, which we are facing in the case of the COVID-19 pandemic. 

Nevertheless, we use two real contact networks to study the impact of the same isolation strategies. We include the Lyon school network ($LS$) with $N=242$ nodes \cite{genois2018can}, and a US high school network ($US$) with $N=788$ nodes \cite{salathe2010high}. Both networks have been constructed using proximity data (in the range of 1-3m) gathered from sensors worn by school staff and students. 

\begin{figure}[!hbtp]
\centering
\includegraphics[width=1.0\linewidth]{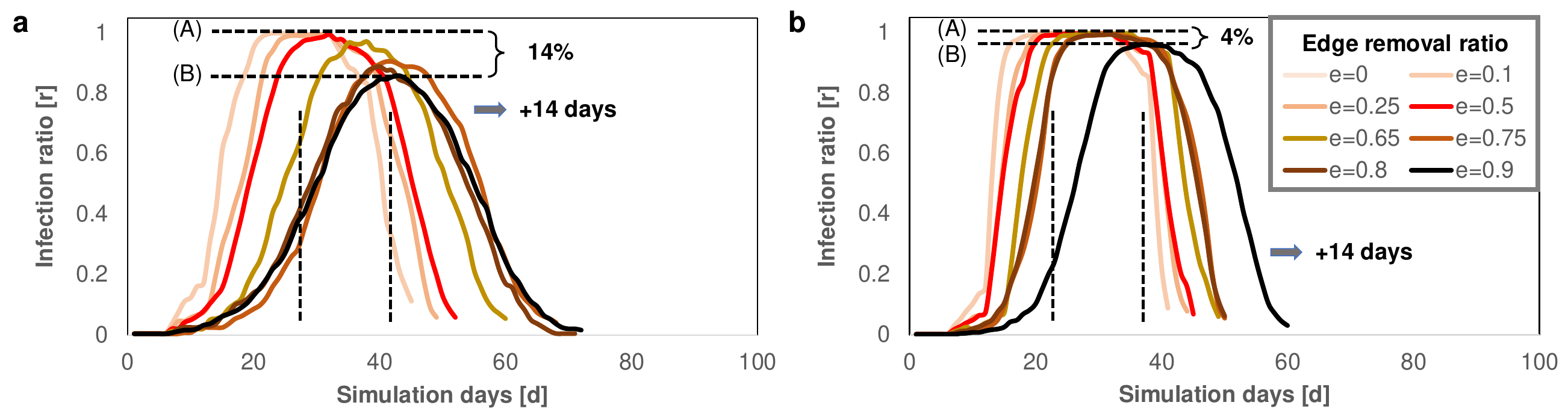}
\caption{The impact of the edge removal ratio $e$ in reducing and delaying the maximum outbreak infection ratio $r$, by applying a combined (C+D) isolation strategy on two micro-scale contact networks: Lyon school \textbf{(a)} and US high school \textbf{(b)}. On both networks, mild to moderate isolation ($e=0-0.5$) has no effect in limiting the infectious spreading (points A); only more severe isolation  ($e>0.75$) manages to reduce the peak of the infection ratio, but only by 4-14\% ($e=0.9$; points B). The outbreak duration lengthens by roughly 14 days by increasing $e$.}
\label{fig:microscale}
\end{figure}

We find that for such small communities the effect of all isolation strategies is less obvious. The most significant reduction in infection ratio $r$ is achieved, again, by the hybrid (C+D) strategy, which we depict in Figure \ref{fig:microscale}a ($LS$ network), and Figure \ref{fig:microscale}b ($US$ network). On both networks, the mild to moderate isolation ($e=0-0.5$) plays almost no role in reducing the infectious spreading (points A), with $r \approx 100\%$. A reduction of the peak infection ratio is only visible for the severe isolation  ($e>0.65$); in case of the maximum isolation with $e=0.9$, the infection ratio drops by 14\% ($LS$) and 4\% ($US$) (point B). In contrast, the infection reduction on the large scale test networks, for the same isolation strategy, is 98\%, compared to no isolation at all, similar to Singapore \cite{koo2020interventions}. Furthermore, the outbreak duration on contact networks $LS$ and $US$ is lengthened by about 14 days when increasing the edge removal ratio from $e=0$ to $e=0.9$.

\section{Discussion}

According to our SICARS epidemic model, we find that a trade-of exists between the the COVID-19 outbreak duration and infection rate. Specifically, applying forms of moderate to strict quarantine can prolong the pandemic duration by roughly +40\% (Hybrid C+D) and +80\% (C) days, compared no isolation at all, where we estimate the duration of the pandemic at $\approx 100$ days. Indeed, while a more extended quarantine may have complex economic consequences, the invaluable benefit of applying isolation strategies is the significant reduction of the peak infection ratio, and implicitly a reduction of casualties.

\begin{figure}[!hbtp]
\centering
\includegraphics[width=1.0\linewidth]{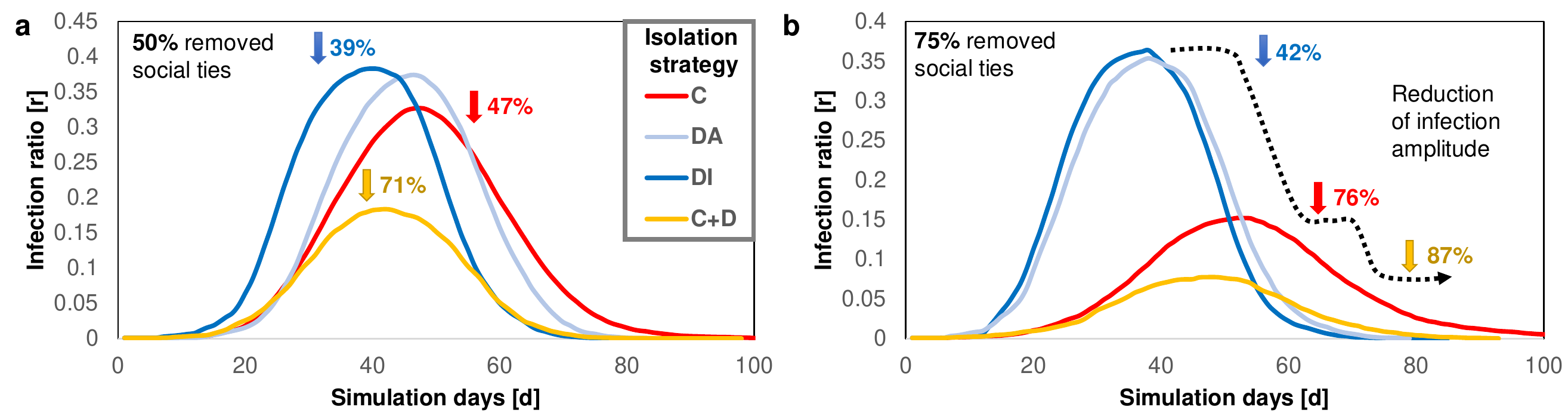}
\caption{Efficiency of isolation strategies, in terms of the relative infection ratio reduction, compared to the baseline case of no isolation (given as percentage next to each corresponding isolation strategy) when the isolation strategy removes 50\% of social ties from the network (moderate isolation) \textbf{(a)}, 75\% of social ties (severe isolation) \textbf{(b)}.}
\label{fig:compare-c-d-cd}
\end{figure}

The overall difference between the C (centralized), D (decentralized), and hybrid (C+D) isolation strategies is summarized in Figure \ref{fig:compare-c-d-cd}. Panel  \ref{fig:compare-c-d-cd}a highlights the reduced efficiency of the DA and DI isolation when removing 50\% of social ties. Compared to the no isolation case, DA and DI manage to reduce the infection amplitude by only 39\%. In contrast, the efficiency of the C isolation grows roughly linearly with $e$, while the hybrid C+D isolation yields the best results (71\% infection reduction). A more strict restriction might be necessary for COVID-19, meaning a reduction of up to 75\% social ties, as depicted in panel \ref{fig:compare-c-d-cd}b. In this case, the efficiency of the C isolation remains proportional to $e$ (75\%), while the combined C+D isolation offers the most significant infection amplitude reduction of 87\%. Our experimental results suggest that by further restricting the quarantine to $e=0.9$, we can achieve a 98\% reduction in outbreak impact, comparable to the 99.3\% results obtained on the Singapore population \cite{koo2020interventions}.

By analyzing different parameters which characterize the course of the COVID-19 pandemic, we find that: 
\begin{itemize}
\item While an early response to adopting isolation strategies is critical, we find that a delay of $d=20-30$ days has a significant negative impact (+8-25\% more infected). Higher delay values of $d=50$ days would have a catastrophic negative impact (+76\%, compared to no delay).
\item While the real patient relapse rate is not yet fully known (\textit{i.e.}, currently is estimated at $<1\%$), we find that a higher relapse rate (\textit{e.g.}, 10\%) would not significantly alter the effectiveness of the isolation strategies, neither in terms of infection amplitude, nor maximum duration.
\item From a network structure standpoint, the C isolation strategy mitigates the superspreaders by immediately reducing the maximum degree in the network (\textit{i.e.}, proportionally to $e$.) The D isolation also reduces the maximum degree significantly (as low as $<10\%$), but only after a delay of $d=36-37$ days.
\item A reactive isolation policy which implies step-wise delayed increases of $e$, is less effective than a proactive policy with a specific $e$ imposed from the outbreak onset. In other words, to match the proactive policy, a reactive policy would have to use a far more severe social distancing ($e$) over time.
\end{itemize}

As a comparison between proactive (or early) and reactive (or delayed) measures applied in real-world settings, we exemplify the cases of China, Singapore, or South Korea \cite{kupferschmidt2020china,lai2020severe} for the former type of response, respectively the UK \cite{mahase2020covid} or USA \cite{saad2020spectrum} for the latter. Indeed, the proactive policy confirms its significant effectiveness, as both the number of infected cases and deaths are already declining in the Asian countries mentioned above. In contrast, current estimations of the COVID-19 progress in Western countries are less optimistic.

There are several limitations to our analysis. On the one hand, based on current evidence, we use biological parameters that characterize the SARS-CoV-2 virus that are not fully confirmed in practice. For instance, we have explored the unknown susceptibility rate (relapse probability) and found that its effects on our conclusions are not major; however, one may update these values as more comprehensive data surfaces. 
On the other hand, the entire contact networks of large regions around the world affected by COVID-19 (\textit{e.g.}, China, Europe, or the USA) are impossible to reconstruct entirely due to their size, dynamics, and lack of individual-level information.
Similar to other approaches \cite{pastor2015epidemic,viboud2018rapidd,pastor2002immunization}, we base our study on several synthetic and real-world datasets as reliable proxies of the contact network of individuals.
By fitting the SICARS model to multiple complex network topologies, we aim at making the best possible assessments on the SARS-CoV-2 transmission dynamics.


Also, our study on the impact of isolation strategies extended to micro-scale contact networks shows several interesting insights. First, due to the small scale and high connectivity of these contact networks, the isolation strategies have little efficiency in defending against the COVID-19 spreading. We believe that the specific duration from contact to symptom onset, characterizing the SARS-CoV-2 virus, renders any measures almost useless in such small communities. The maximum possible reduction of infection ratio of 4-14\% is obtained only for the most severe isolation ($e=0.9$). Second, we observe a lengthening of the outbreak duration (by 14 days), similar to the one visible on the large scale test networks. Third, we conclude that the topologies of contact networks are representative only for micro-scale communities (\textit{e.g.}, the Diamond Princess cruise ship quarantined in Japan, or the Theodore Roosevelt aircraft carrier docked in Guam), and may not be directly extrapolated at the level of large cities, regions, or countries, as is the case of the COVID-19 pandemic.

Of note, our observations come in support of a recent scandal in the U.S. navy involving the decision to remove infected military personnel from an aircraft carrier. Indeed, due to the high density of the contact network aboard, and the limited physical space making social distancing almost impossible, our results suggest that the captain took the right decision. Indeed, without such measures the SARS-CoV-2 virus spreading would have been super fast and may have likely achieved an infection ratio of up to 100\% in no time.

Summarizing our simulation results, the main conclusions of this study are:

\begin{itemize}
\item The decentralized (D) isolation strategy helps\footnote{As it can be considered a characteristic of antifragile societies that have recently experienced large-scale emergency situations such as SARS, tsunamis, earthquakes \cite{taleb2012antifragile,topirceanu2020complex} \textit{etc}.}, but is \textit{insufficient} on its own. In other words, the decentralized methods are not effective in mitigating the impact of the superspreaders (\textit{i.e.}, nodes with a high connectivity, therefore a high contagious potential) in a timely manner; hence, their potential for reducing the dynamics of pandemics like COVID-19 is limited.
\item The centralized (C) isolation strategy is a must, as it is more efficient than the decentralized approaches.
\item The best response strategy is a hybrid one (C+D). 
If used too late after the outbreak of the pandemic, even the combined C+D strategy loses its effectiveness significantly.
\item All isolation strategies applied \textit {proactively} (at outbreak onset) are more effective than applied \textit{reactively} (in response to the epidemic dynamics) even by progressively increasing the isolation severity.
\item A higher patient relapse rate than currently estimated for COVID-19 would not alter our conclusions on the effectiveness of the isolation strategies significantly.
\end{itemize}

With COVID-19 having reached the status of a global pandemic, understanding the effectiveness of social distancing measures in various settings is crucial for apprehending the dynamics of the outbreak and being able to contain or mitigate it adequately.


\bibliographystyle{unsrt}  
\bibliography{bibliography} 
\end{document}